\documentclass[10pt,twocolumn]{article}
\usepackage{times}
\usepackage{amsmath}
\usepackage{amsfonts}
\usepackage{amssymb}
\usepackage{graphicx}
\usepackage{wrapfig}
\usepackage[margin=0.75in]{geometry}

\usepackage[font=footnotesize,labelfont=bf]{caption}
\usepackage{chngcntr}
\counterwithout{figure}{section}

\usepackage{verbatim}
\usepackage{hyperref}

\usepackage[numbers, comma, sort&compress]{natbib}
\usepackage{sidecap}

\usepackage{color}

\usepackage{natbib}
\usepackage{hyperref} 
\title{Viewing Earth's surface as a soft matter landscape}

\usepackage{authblk}

\author[1,2]{Douglas J. Jerolmack}
\author[3]{Karen E. Daniels}

\affil[1]{\small Department of Earth and Environmental Science, University of Pennsylvania. Phone: 215-746-2823. Fax: 215-898-0964. sediment@sas.upenn.edu}
\affil[2]{\small Department of Mechanical Engineering and Applied Mechanics, University of Pennsylvania}
\affil[3]{\small Department of Physics, North Carolina State University. Phone: 919-513-7921. kdaniel@ncsu.edu}

\date{\today}

\begin{document}

\maketitle

\noindent
\textbf{Abstract}: The Earth's surface is composed of a staggering diversity of particulate-fluid mixtures: dry to wet, dilute to dense, colloidal to granular, and attractive to repulsive particles. This material variety is matched by the range of relevant stresses and strain rates, from laminar to turbulent flows, and steady to intermittent forcing, leading to anything from rapid and catastrophic landslides to the slow relaxation of soil and rocks over geologic timescales. Geophysical flows sculpt landscapes, but also threaten human lives and infrastructure. From a physics point of view, virtually all Earth and planetary landscapes are composed of soft matter, in the sense they are both deformable and sensitive to collective effects. Geophysical materials, however, often involve compositions and flow geometries that have not yet been examined in physics. In this review we explore how a soft-matter perspective has helped to illuminate, and even predict, the rich dynamics of Earth materials and their associated landscapes. We also highlight some novel phenomena of geophysical flows that challenge, and will hopefully inspire, more fundamental work in soft matter.

\bigskip


\section{Introduction}

\begin{figure*}
    \centering
    \includegraphics[width=\linewidth]{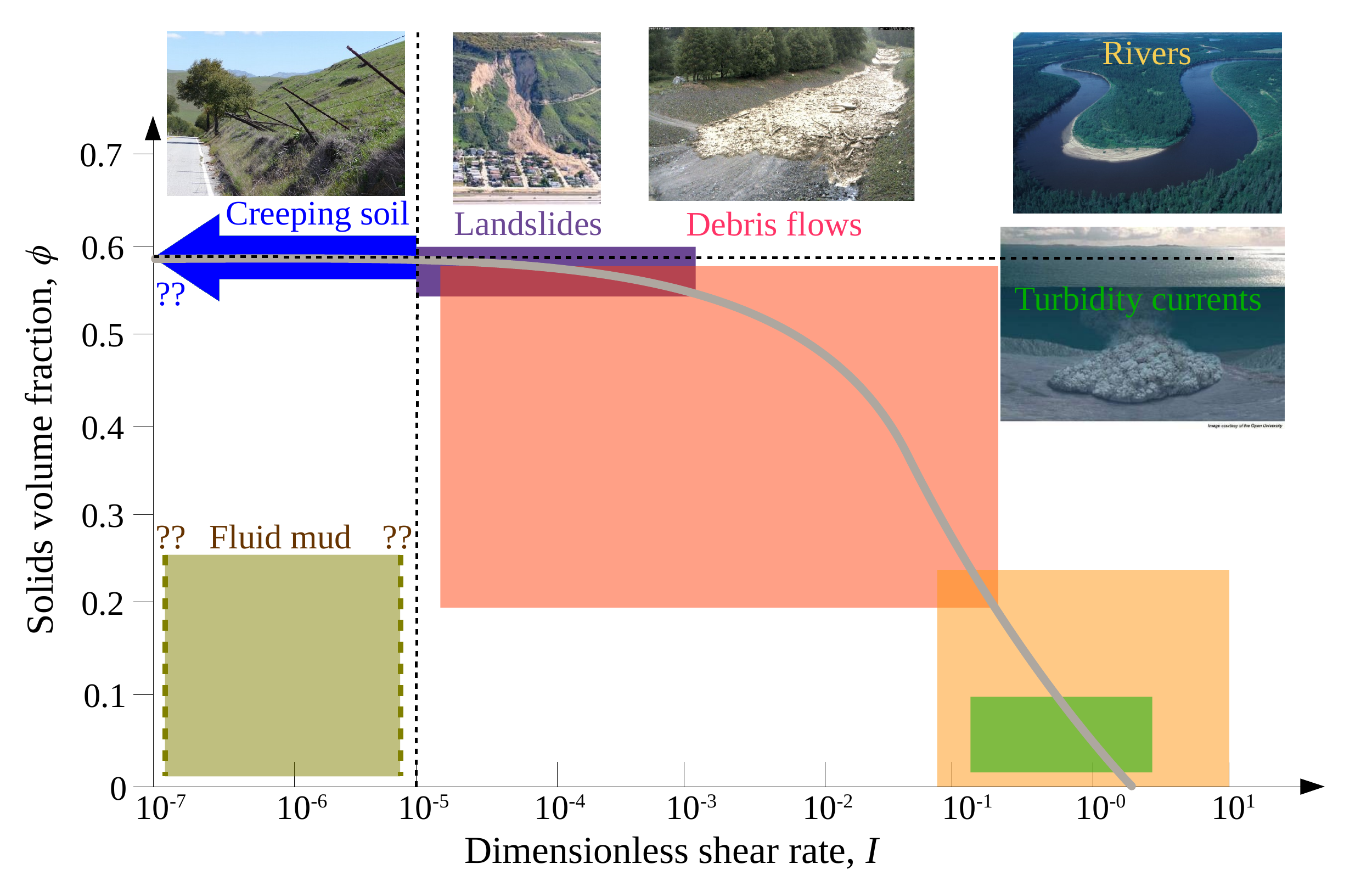}
    \caption{Phase diagram of particulate geophysical flows discussed in this paper. Each colored box marks the typical range of parameters in which a particular type of \textit{bulk flow} is observed, based on values reported in the literature and reasonable estimates of confining pressure. Boundary values for $I$ for fluid muds are completely unknown and in-situ images are unavailable. The two black dashed lines mark the approximate boundaries of the solid-fluid transition; the yielding line for $I \sim 10^{-5}$ determined from \cite{houssais2016rheology, ferdowsi2018glassy}, and $\phi = \phi_c \approx 0.59$ chosen from experiments of \cite{boyer2011unifying}. The solid grey line suggests a $\mu(I)$-type relation (see Box 1) to guide the eye; it follows the data from experiments on fluid-sheared sedimenting particles by \cite{houssais2016rheology}. Note that fluid muds, in which cohesive/attractive particle interactions are significant, do not fall on the grey line --- indicating major deviations from $\mu(I)$ rheology.
   Image credits ---  
   Creeping soil: \url{https://mapio.net/pic/p-70357329}; Landslide: Photograph by Mark Reid, US Geological Survey;
   Debris flows: \url{http://www.irpi.cnr.it/en/focus/debris-flow-monitoring/};
   Rivers: US Fish and Wildlife Service, accessed through \url{https://serc.carleton.edu/eyesinthesky2/index.html};
   Turbidity currents: \url{https://www.wired.com/2011/11/ideas-about-the-origin-of-submarine-canyons-from-the-1930s/}.
    \label{fig:phase}}
\end{figure*}

Patterns on the Earth's surface are created by geophysical flows, composed of  fluid-particle mixtures of varying proportions from dry to wet to immersed \cite{Bagnold1941, Bagnold1954, seminara2010fluvial, charru2013sand} (Fig. \ref{fig:phase}). These patterns form landscapes that provide the template for human settlement, but their unpredictable dynamics also create natural hazards that threaten lives and infrastructure \cite{huppert_extreme_2006, syvitski2005impact}. Familiar features such as canyons, sand ripples, dunes, river channels, and deltas also form in the deep ocean \cite{canals2004slope, rebesco2008contourites} and are ubiquitous in the solar system \cite{greeley2013introduction}. This similarity, despite the exotic nature of some fluid and solid materials involved (e.g., liquid methane and water-ice particles on Titan), both motivates and challenges our understanding of the underlying physics \cite{grotzinger2013sedimentary}. This ``the science of scenery'' \cite{nature1928science} is called geomorphology, and a central challenge is to understand and link the mechanics of geophysical flows to the evolution of landscapes that results from the cumulative effects of innumerable flow events \cite{anderson2010mechanics}. 

Remarkably, the Earth is  ``soft'' on geological timescales if we take the meaning of that term in the spirit of de Gennes \cite{de1992soft}: our ground is composed of materials that are responsive in their collective effects. However, only the patient observer will notice the relaxation of mountains as rocks flow at speeds of 0.1~nm/s ($10^{-2}$~m/yr). Slightly faster are the rates at which soil and ice creep downhill, sometimes exceeding 100~nm/s ($10$~m/yr). Yet, true to the typical sensitivity of soft materials, these processes can intermittently become unstable and landslides can reach speeds of 10~m/s (Fig. \ref{fig:phase}). 

Furthermore, earth materials exhibit such soft-matter effects as shear-rate dependent rheologies influenced by microstructure \cite{kang2012simulation, houssais2016rheology, vasisht_rate_2018, ghosh_direct_2017}; aging and history dependence \cite{courtland_direct_2002,bonn_laponite:_2002, hartley_logarithmic_2003, charru2004erosion, turowski2011start, masteller2017interplay, bililign_protocol_2019}; and signatures of glassy dynamics and jamming \cite{weeks_three-dimensional_2000,keys_measurement_2007,frey2011bedload, charbonneau_glass_2017, ferdowsi2018glassy}. As such, the same underlying causes that have engaged soft condensed-matter physicists --- excluded-volume effects, the emergence of bulk properties such as rigidity from particle-scale interactions, and the role of disorder in dynamical phase transitions  --- are also at play in sculpting the landscapes we see around us. 

That said, geophysical flows are far from the idealized granular flows and suspensions usually considered in physics (Fig. \ref{fig:phase}). Particulate earth materials typically have strong heterogeneity in grain size and composition, are often cohesive, encompass a vast range of pressures and timescales, and are subject to time- and space-varying forcing. Perhaps most challenging and intriguing is that geophysical flows make their own boundaries; landscape patterns are an expression of the competition between forcing at the {\itshape interface} (rainfall, wind and water currents, uplift of tectonic plates, gravity) and rheology in the {\itshape bulk}.

In spite of these differences, recent experimental advances in the field are illustrating how
emerging unifying concepts in soft matter can be meaningfully applied to describe natural landscapes, and also how landscapes can present novel materials and experimental configurations that may challenge and illuminate the basic physics. In this review, we will focus our attention on particulate systems, where lessons from granular materials and suspensions translate most directly; nonetheless, clear connections also exist to other amorphous earth materials such as rock and ice.
For many of these topics, there are prior reviews of key soft-matter concepts and techniques that we will lean on: rheology and yielding in soft materials \cite{chen_rheology_2010, falk2011deformation, denn2014rheology, bonn2017yield, guazzelli2018rheology, nicolas_deformation_2018} and their application to geophysical flows \cite{houssais2017toward}; jamming \cite{liu_jamming_2010} and glassy dynamics \cite{falk2011deformation, charbonneau_glass_2017,nicolas_deformation_2018}; granular segregation \cite{gray_particle_2018}; and a variety of experimental soft-matter \cite{NagelSoft2017} and granular \cite{amon_preface:_2017} techniques.

The application of granular physics to understanding fault dynamics and earthquakes is well established \cite{marone1998laboratory, daniels_force_2008, hayman_granular_2011, elst2012auto, ferdowsi2015acoustically,  nicolas_deformation_2018}. The importance of granular contributions to geomorphology, however, is just starting to gain attention \cite{frey2011bedload, houssais2017toward}. The dominant framework for describing particulate (sediment) transport has been fluid mechanics, and for good reason. Water and wind form turbulent boundary-layer flows on the Earth's surface, which produce time- and space-varying stresses that entrain and suspend particles \cite{parker1987experiments, seminara2010fluvial, martin2017wind}. In addition, high-concentration particulate flows such as landslides can be described as viscoplastic fluids \cite{iverson1997debris, iverson_flow_2001, houssais2017toward}. The scope of our paper is framed by this context, as well as studies presenting successful complementary views on geomorphology, most notably: the formal statistical mechanics formulation of sediment transport by Furbish and colleagues \cite{furbish2012probabilistic, furbish2017elements}, and related probabilistic and stochastic approaches in geomorphology \cite{einstein1950bed, dodds2000scaling, schumer2009fractional, ancey2015stochastic}; nonlinear/dynamical-systems approaches to landscape pattern formation \cite{rodriguez2001fractal, murray2009geomorphology, devauchelle2012ramification, goehring_pattern_2013}; hydrodynamic and classical stability analyses \cite{seminara2010fluvial, charru2013sand}; and reviews justifying the applicability of small-scale experiments to natural landscapes \cite{paola2009unreasonable, malverti2008small}. 

What makes soft matter distinct from these other approaches is its primary focus on the properties and behavior of disordered materials. Our review emphasizes that the central questions in soft-matter physics today are also central questions in geophysical flows: the effects of polydispersity in particle size and shape, particle attraction, memory/aging, and mechanical perturbations/excitations, on state transitions and rheology.

\section{Classification of geophysical flows}

In Figure~\ref{fig:phase} we introduce a range of particulate geophysical flows that are discussed in this paper, in the phase space of two parameters: the dimensionless strain rate $I \equiv {\dot \gamma} t_\mathrm{micro}$, where ${\dot \gamma}$ is strain rate and $t_\mathrm{micro}$ is a microscopic timescale of particle motion \cite{midi2004dense, boyer2011unifying, guazzelli2018rheology}; and volume fraction $\phi$ (see Box 1). {\it Soil creep} is the sub-yield, quasi-static, downslope motion that occurs on hillsides \cite{bishop1960factors, culling1963soil, garlanger1972consolidation, roering2004soil} which corresponds to the largest $\phi$ values and smallest $I$ values; its lower bound is unknown due to measurement limitations, but the arrow is meant to indicate that it is many orders of magnitude smaller than the range shown \cite{ferdowsi2018glassy}. {\it Landslides} are dry to partially-wet, dense granular flows that move down hillsides; the lower $\phi$, as compared to creeping soil, arises because shear results in dilation of the flow \cite{iverson1997debris, okura2002landslide, iverson_landslide_2015}. {\it Debris flows} are dense granular suspensions, often formed from progressive wetting of landslides, that typically move down river channels \cite{iverson1997debris, iverson_flow_2001, iverson2010perfect}. {\it River flows} have average particulate concentrations that are typically in the very dilute suspension regime, but may reach moderate density suspensions in some cases. Note, however, that rivers drive \textit{interfacial} flows of dense-granular and creep regimes at their boundaries (Fig.~\ref{fig:classify}; see below). {\it Turbidity currents} are particulate suspensions whose buoyancy contrast is sufficient to drive flow, but $\phi$ is low enough that turbulence is sufficient to keep the grains suspended \cite{kuenen_turbidity_1950, meiburg2010turbidity}. They typically form in the ocean due to the collapse of granular material and subsequent (turbulent) entrainment of ambient water \cite{kuenen_turbidity_1950, parker1987experiments, meiburg2010turbidity, you2012dynamics}. Because they occur on the seafloor, typically under kilometers of water, an artist's rendering is provided as an illustration. {\it Fluid muds} are dilute to moderately dense, quasi-stable ``colloidal gels'' \cite{colombo2014stress, bonn_yield_2017} that form in estuaries and coasts from cohesive clays and organic materials delivered by rivers \cite{winterwerp2002flocculation, mcanally2007management}. They are distinct from the other flows shown due to the small particle sizes and associated inter-particle attraction, which is induced when particles enter salty water and the typically-repulsive surface charges are screened by dissolved ions \cite{coussot1994behavior, winterwerp2002flocculation}. Although the bounding values of $I$ are unknown for such flows (question marks), attractive particle interactions allow a yield stress to develop at lower-than-expected $\phi$ values \cite{mcanally2007management}. 

Deformation of soft materials is exquisitely sensitive to the nature of the forcing, including boundary conditions.
Geophysical flows may be usefully placed into two broad categories based on their dynamics (Fig. \ref{fig:classify}). In an {\it interface-driven flow}, energy transfer at the interface between two materials with contrasting densities causes the shape of the interface to evolve in both space and time. These driven dynamics can give rise to characteristic interface shapes, including well-defined wavelengths, as is common in non-equilibrium systems \cite{cross_pattern_1993}. In some cases, the inertia of moving fluid can push the interface into a new shape, as occurs for ripples on the surface of cooling lava \cite{griffiths_dynamics_2000}. The more common situation on the Earth's surface is fluid-driven sediment transport, such as: wind-blown sand on the surface of dunes \cite{Bagnold1941,howard_sand_1978}, underwater sand ripples and dunes \cite{ayrton_origin_1905, charru2013sand}, turbidity currents \cite{kuenen_turbidity_1950, meiburg2010turbidity}, and meandering river channels \cite{einstein_ursache_1926, seminara2010fluvial}, all of which involve both erosion and re-deposition of particulate material along the interface. Modeling efforts for such systems have traditionally focused on describing the evolution of the interface's shape, rather than the bulk material underneath \cite{seminara2010fluvial}. Mechanics-based formulations for sediment transport envision a thin film of particles, with momentum supplied from the driving fluid, transported over a static underyling bed \cite{raudkivi1998loose, charru2004erosion, lajeunesse2010bed}. The separation between this flowing {\it bed-load} layer and the substrate, however, is not so sharp; interfacial grain motion bleeds downward into the bulk due to granular shear, inducing creep deep beneath the bed \citep{Houssais2015, maurin2016dense, allen2018granular} (Fig. \ref{fig:classify}a). Thus the fluid-particle interface is fuzzy, and it grows or contracts with changes in the applied fluid stress \cite{capart2011transport, aussillous2013investigation, Houssais2015}; nevertheless, particulate transport is boundary driven.    

This is in contrast with Earth materials undergoing {\it bulk flow}, in which the interface changes shape as a consequence of deformations taking place throughout the material. In nature, these flows are typically gravity (rather than fluid) driven. Common examples of non-inertial bulk flow are creeping soil on hillslopes \cite{roering2004soil,ferdowsi2018glassy} (Fig. \ref{fig:classify}b), slumping of the continental shelf \cite{laughton_morphology_1978}, viscous relaxation of volcanoes \cite{byrne_sagging-spreading_2013}, or slowly flowing glaciers \cite{goldsby2001superplastic}. In these, the material itself can be considered to have enough cohesion and internal rigidity to behave as a solid on seasonal timescales, yet appears to flow slowly on geological timescales. Rapid bulk flows also occur, when inertial forces propagate all the way through a material. Examples include soil liquefaction \cite{ishihara_liquefaction_1993}, and fluidization in debris \cite{iverson1997debris} and pyroclastic \cite{breard2016coupling} flows (Fig. \ref{fig:phase}). In these systems, the volume of the particulate flow --- and hence the location of the particle-fluid (soil-air) interface --- is determined in part by shear-induced dilation and pore-pressure effects within the bulk \cite{iverson1997debris, iverson_flow_2001, iverson2010perfect}.

These two types of mechanisms can of course both be present. One example is sedimentation, which can be thought of as the emergence of a boundary between the solid and liquid phase. The interface is formed as the settling of particles is hindered by crowding, while the downward speed of the interface is determined by further consolidation of the bulk \cite{guazzelli2011physical}. As expected for soft materials, sedimentation dynamics are strongly influenced by thermal \cite{ortiz_flow-driven_2013,brzinski2018observation} effects and interparticle attraction/repulsion \cite{brzinski2018observation, sutherland2015clay}. One example is turbidity currents: the upper boundary forms by sedimentation; however, this interface evolves due to both mixing/settling within the current, and de-stabilizing fluid shear at the boundary \cite{kuenen_turbidity_1950, parker1987experiments, meiburg2010turbidity, you2012dynamics} (Fig. \ref{fig:geo_materials}). Fluid muds are a similar case: sedimentation occurs as attractive particles aggregate, but aggregates are resuspended by waves and currents \cite{mcanally2007management}. Besides sedimentation, other examples of combined interfacial and bulk deformation may be found in landslides and glaciers, where shear localization at the base of these flows produces a lower interface between the bulk flow and the underlying substrate (Fig. \ref{fig:classify}c). This lower boundary may be bedrock or an internal slip plane; in either case shear-banding may occur at the interface, often due to large confining pressures and the associated effects of lubrication/pore pressure \cite{clarke1987fast, okura2002landslide}.

\begin{figure}
\includegraphics[width=\linewidth]{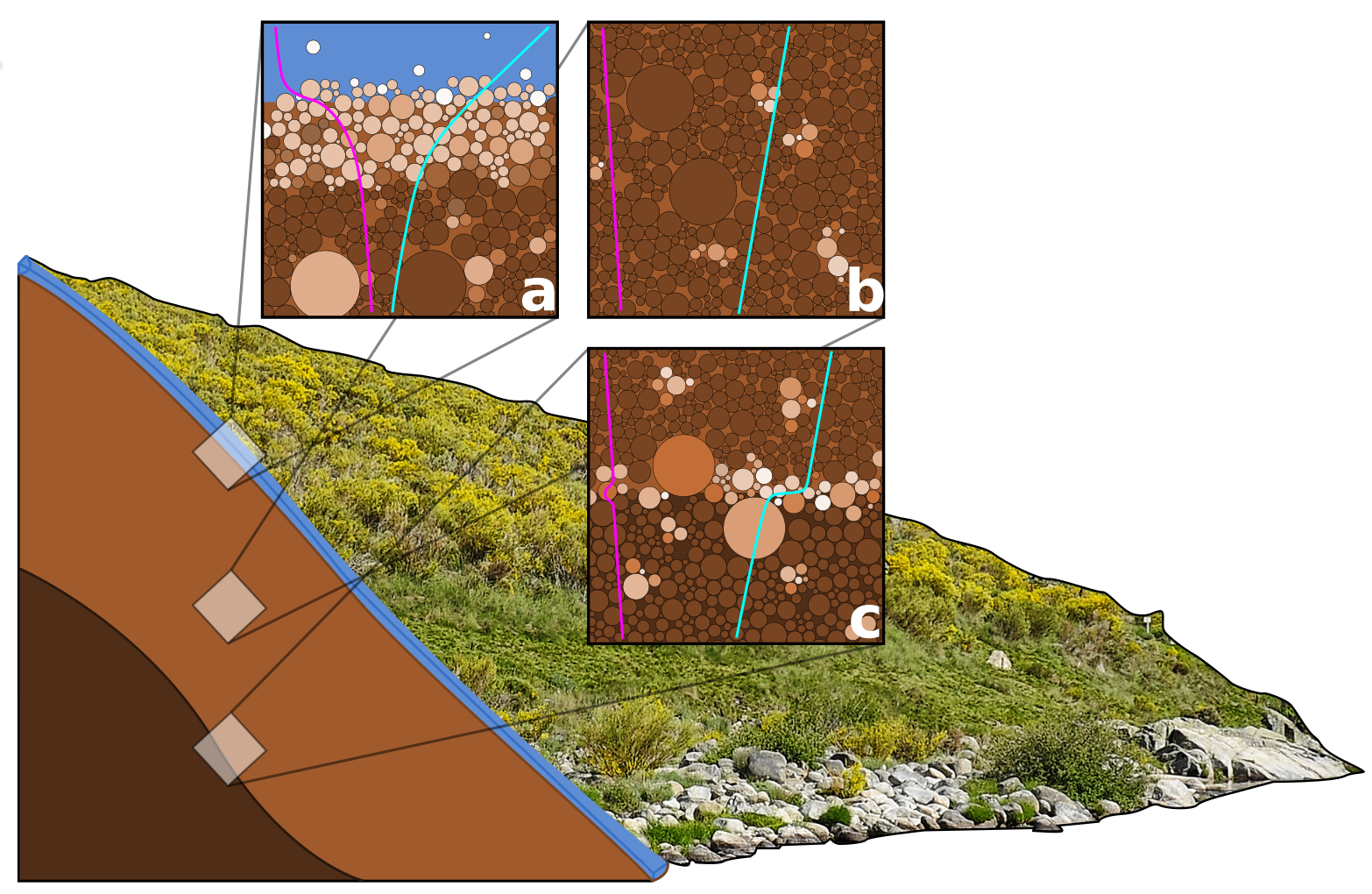}
\caption{Interfacial and bulk dynamics of geophysical flows, illustrated on a prototypical soil-covered hillslope. Each inset has schematic log-linear profiles of particulate volume fraction $\phi$ (magenta) and down-slope velocity $u$ (cyan) overlaid; grain color is relative speed, scaled from mobile (white) to immobile (brown). (a) Blue line represents a small river, an example of an \textit{interface-driven flow} where fluid shear from above drives deformation of the particle-fluid interface; granular shear drives motion deep into the substrate, which transitions at depth to creep. Generalized from \cite{Houssais2015, houssais2016rheology}; see text for more explanation. (b) Gravity-driven \textit{bulk deformation} of the polydisperse granular soil; example shows soil creep, which is accommodated by local and rare rearrangements as the landscape relaxes in response to disturbance. Both $u$ and $\phi$ are typically observed to decrease exponentially with depth due to granular friction and compaction, respectively. (c) A shearing interface accommodated by dilation appears where local rearrangements critically percolate to facilitate slip; this is often the base of catastrophic failure. Schematic after experiments by Amon et al. \cite{amon2013experimental}.}
    \label{fig:classify}
\end{figure}

\section{Soft matter concepts in earth materials}

\paragraph{Rheology:} Soft-matter physics and geomorphology are long-lost relatives, as they share important components of their origins in the pioneering work of Bagnold. He recognized that geophysical flows span a gradient from granular to hydrodynamic control --- what we would today call dense-granular flows to dilute suspensions --- and sought a generalized rheology to connect the grain-inertia to fluid-viscosity dominated regimes \cite{Bagnold1954}. This class of materials are now referred to as {\it granular suspensions } \cite{guazzelli2018rheology}, mixtures of fluids with non-Brownian and non-attractive particles. Key insights of {\it Bagnold rheology} are that dissipation via collisions depends on both particle concentration and shear rate (Box 1). This work formed the foundation for Bagnold's approach to sediment transport in rivers \cite{bagnold1956flow, hunt2002revisiting}, and a kernel of it survives in the constitutive relations employed in geophysical flows today \cite{iverson2010perfect, capart2011transport}. Bagnold rheology was the seed for the so-called $\mu (I)$ {\it rheology,} a phenomenological constitutive relation between an effective friction ($\mu$) and the dimensionless shear rate ($I$) that has recently been shown to unite granular and suspension rheology \cite{boyer2011unifying, guazzelli2018rheology} (Box 1). 

\begin{figure*}
\centering
\includegraphics[width=0.9\linewidth]{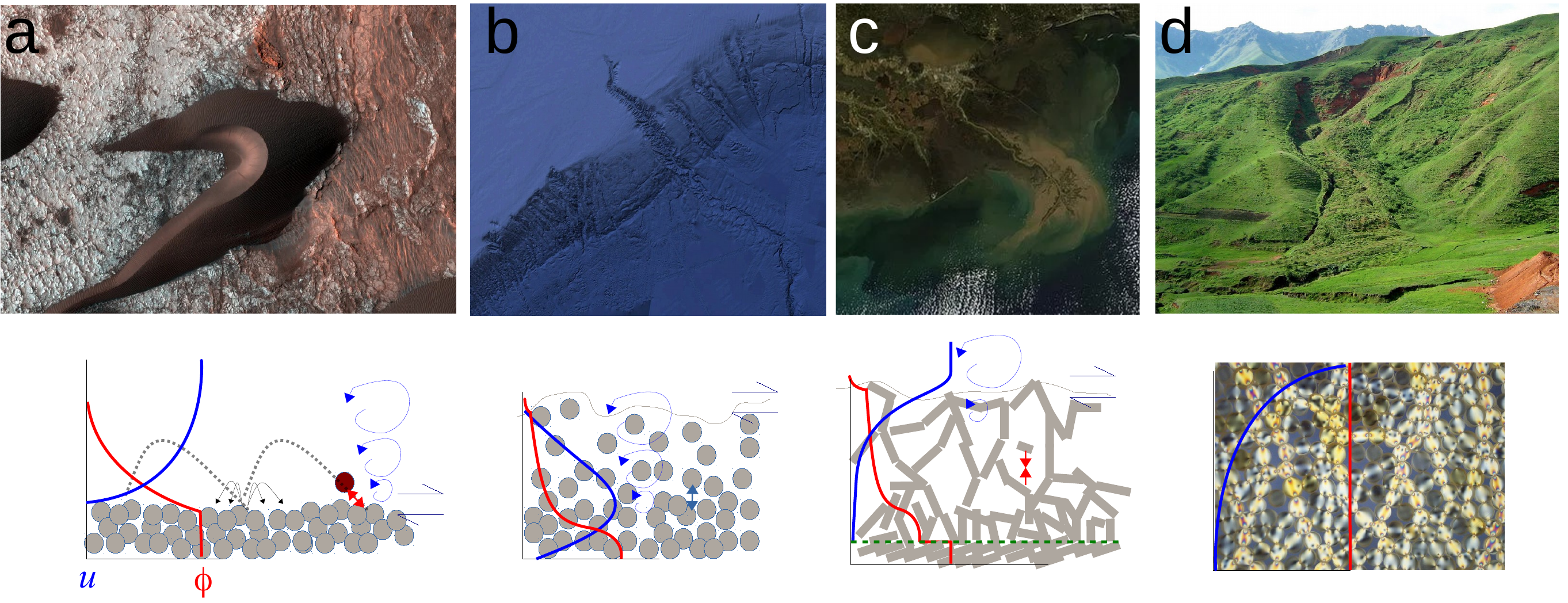}
\caption{Example landscape patterns (top), and particle and fluid interactions of associated flows (bottom). Blue and red curves correspond to particle/fluid velocity ($u$) and particle volume fraction ($\phi$) profiles, respectively; blue swirls indicate where turbulence is relevant, and opposing half arrows indicate fluid shear at an interface. (a) Sand dune on Mars (image credit: NASA/JPL) created by wind-driven sand transport (https://doi.org/10.1017/jfm.2015.601). Particle interactions are partially elastic (repulsive) collisions of grains with the bed (dashed lines show trajectories), which can {\it splash} up other grains (black arrows). (b) Submarine canyon and channel carved by turbidity currents (image credit: USGS). Particles within current interact via hydrodynamic repulsion and lubrication, forming a sheared hindered settling interface (traced in grey), and turbulence keeps grains suspended. (c) Mud suspension emanating from the Mississippi Delta (image credit: NASA). Some clay-rich suspensions form fluid muds by particle attraction, which is sheared from above (grey line). Green dashed line marks boundary with irreversibly compacted mud substrate below. (d) A creeping landslide or earthflow (image credit: Wikipedia). Highly concentrated particles interact frictionally under gravity-driven, quasi-static shear; force chains become important for stress transmission, where color intensity corresponds to force magnitude (Source: KED) \cite{majmudar_contact_2005}.  \label{fig:geo_materials}}
\end{figure*}

Earth materials that fall into this category of granular flows/suspensions include landslides \cite{iverson_landslide_2015, ancey2001role}, debris flows \cite{turnbull_debris_2015, iverson1997debris}, and river sediment transport \cite{houssais2016rheology, maurin2016dense} (see Fig.~\ref{fig:phase}). They are characterized by a viscoplastic rheology; materials are solid(-like) below a critical shear rate (or yield stress) associated with inter-granular friction, and exhibit shear-rate dependent viscosity (or friction) above yield. Two primary challenges to rheological descriptions of these particulate systems are: (1) accounting for sub-yield creep, which is ubiquitous in granular materials \cite{reddy2011evidence, falk2011deformation, bandi2013fragility, Houssais2015, amon2013experimental, pons2016spatial, allen2018granular, ferdowsi2018glassy}; and (2) how to correctly couple rheological models to the boundaries. Recently, nonlocal constitutive relations have been proposed that successfully explain the extension of above-yield flow into sub-yield regions for many flow configurations \cite{pouliquen_non-local_2009,kamrin_nonlocal_2012, bouzid_non-local_2015, tang_nonlocal_2018}. Such models cannot, however, account for purely sub-yield creep that occurs even in the absence of any flowing layer \cite{amon2013experimental, allen2018granular}, slip near solid boundaries, the geometry of shear bands \cite{midi2004dense,cheng_three-dimensional_2006}, the transition from inertial to creeping flow \cite{koval_annular_2009}, or fluidization at distances far from disturbance \cite{nichol_flow-induced_2010,reddy2011evidence}. For materials composed of colloidal (rather than granular) materials, additional classes of behavior can arise due to thermal effects or inter-particle attraction \cite{chen_rheology_2010}; each of these classes may be mapped to earth materials in nature (Box 1; Fig.~\ref{fig:geo_materials}).

\paragraph{Rigidity:} The zeroth-order problem in geomorphology is to determine under what conditions material will move, and yet it remains particularly challenging  \cite{Houssais2015}. Typical Mohr-Coulomb failure models do not appropriately describe the solid-liquid phase transition of geophysical flows, whether from entrainment of river-bed sediment by an impinging fluid \cite{frey2011bedload}, or the bulk liquefaction that creates landslides \cite{iverson1997debris, iverson_flow_2001, houssais2017toward}. 
For example, geotechnical models for the latter consider soil to be a solid with a well defined shear strength, above which it yields \cite{Terzaghi1943, schofield1968critical}. Attractive forces between particles (cohesion) effectively raise this shear strength, while increasing pore pressure (due to water content) lowers shear strength by reducing the resisting normal stress \cite{bishop1960factors, gan1988determination}. The slow, sub-yield creeping motion of soil is considered to be a type of viscous flow that is modeled using simple constitutive equations \cite{garlanger1972consolidation, Savage1982}. In principle, the solid-state failure model and the sub-yield creeping `flow' model are physically incompatible. In practice, these models require site-specific calibrations and parameterizations that limit their predictive power \cite{Zieher2017}.

Rigidity transitions of this type are central concepts in soft-matter physics \cite{wyart_rigidity_2005}, and it is possible to draw connections between the frictionless jamming transition \cite{liu_jamming_2010} and rough frictional geophysical flows.
For simplicity here we consider jamming (unjamming) to be a rapid increase (decrease) in rigidity that is typically associated with an increase (decrease) in volume fraction $\phi$ toward (away from) a critical value $\phi_c$. The nature of this transition, however, is sensitive to particle contacts and interactions \cite{basu2014rheology, hsu2018roughness,james2019tuning} and interparticle friction \cite{silbert_jamming_2010,henkes_critical_2010}, factors that are important for geophysically-relevant properties such as dilatancy \cite{schroeter_local_2017}. 

For example, granular materials may jam under shear as force networks are formed, but these states may be fragile or robust depending on the shear-stress magnitude \cite{bi2011jamming, bandi2013fragility} (Fig. \ref{fig:geo_materials}). A related rigidity transition in dense suspensions is {\it discontinuous shear thickening,} which has been suggested to result from a stress-driven transition from lubricated to frictional granular contacts \cite{clavaud2017revealing, morris2018lubricated}. Research has already shown that the solid-liquid transition in geophysical flows is dependent on volume fraction \cite{rondon2011granular, you2012dynamics}, shear stress \cite{iverson1997debris, iverson2010perfect}, and lubrication \cite{mohrig1998hydroplaning, ancey2001role}; concepts from jamming should therefore be readily applicable to Earth materials \cite{houssais2017toward}. Yet, it is unclear whether Earth-surface materials actually jam. The pervasive sub-yield creep observed in granular heaps \cite{amon2013experimental, ferdowsi2018glassy} and fluid-driven granular beds \cite{allen2018granular} occurs at volume fractions and stress values where we might expect athermal materials to be jammed. Yield in these free-surface flows appears to exhibit the dynamics of a {\it glass transition} \cite{amon2013experimental, ferdowsi2018glassy}, where frameworks such as shear transformation zones \cite{falk2011deformation} and depinning \cite{reichhardt2016depinning, aussillous2016scale, ozawa2018random} become relevant; we return to this below. An additional rigidity transition that is relevant for geophysical materials is the {\it sol-gel transition} in attractive colloidal suspensions. Long-range particle interactions may allow the formation of percolated particle networks, and the emergence of an effective yield stress, even at very low $\phi$ \cite{colombo2014stress, bonn2017yield}. These dynamics appear to govern fluid-mud formation \cite{winterwerp2002flocculation, mcanally2007management}.

\paragraph{Excluded volume effects and landscapes of valid states:} A common feature of soft matter systems is the presence of excluded volume effects: a particle/molecule is excluded from accessing some position due to the presence of a particle/molecule at a location overlapping that position. This property is particularly relevant to particulate systems \cite{caglioti_tetris-like_1997}, where the cooperative effects of excluded volume lead to the presence of effective friction \cite{lespiat_jamming_2011,peyneau_frictionless_2008} and cohesion \cite{gravish_entangled_2012}, even in the absence of either material friction or attractive forces. The relative importance of this effect can be determined from the packing fraction $\phi$ on its approach to $\phi_c$ (Fig. \ref{fig:phase}), and the phenomenon is at the heart of many of the jamming/rheological phenomena discussed in Box 1. 

\begin{figure}
\centering
\includegraphics[width=\linewidth]{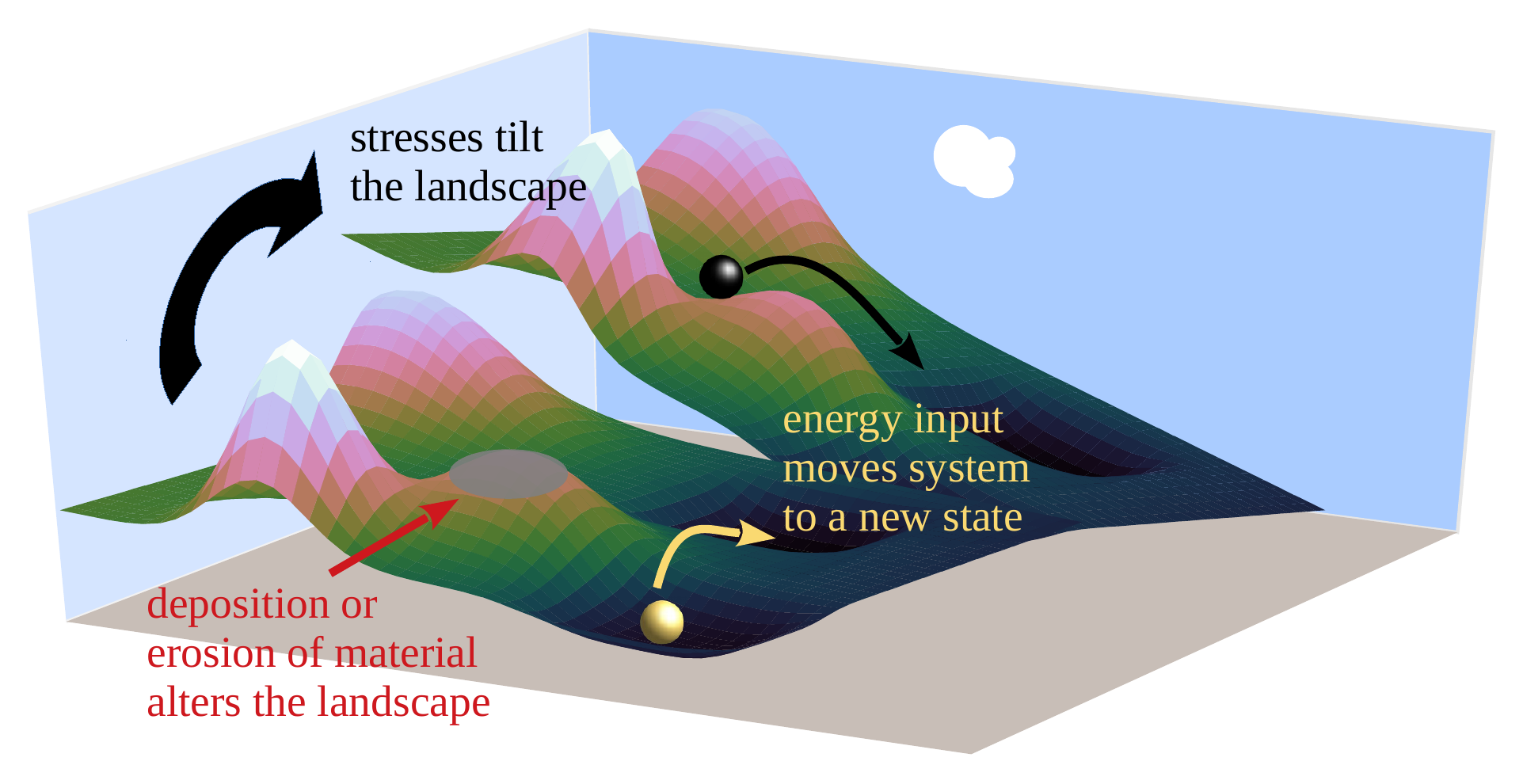}
\caption{The landscape of valid states (landscapes). The balls represent the current state of the system, and must lie on the surface of this landscape of valid states.  External forces can drive the system to find new states. }
    \label{fig:landscape}
\end{figure}

Geophysical flows with particles packed closely enough for these effects to be relevant --- i.e., $\phi \rightarrow \phi_c$ --- are very common (Fig. \ref{fig:phase}).
A consequence of volume exclusion is that certain states are inaccessible within something analogous to the complex energy landscape \cite{wales_energy_2003}, if creating them would require two particles/molecules to overlap. Because granular materials are athermal and non-equilibrium, determining the correct constraints on their states is an open question, and likely involves both positions (volume-exclusion) and stresses (force and torque balance) \cite{bi_statistical_2015}. Taking the analogy to energy landscapes, disordered packings exist in a high-dimensional landscape of valid states. Rearrangements into valid nearby states may be forbidden or favored under some given driving (Fig. \ref{fig:landscape}), and getting trapped for long times in a metastable state is a common occurrence. However, because the Earth is constantly driven, the \textit{real} landscape eventually finds an unstable manifold within the \textit{valid state landscape}, and the dynamics occur along that unstable direction. Therefore, these complex landscapes contribute to the fragile and/or aging nature of many soft materials, and can lead to interesting effects such as metastability \cite{iikawa_sensitivity_2016}, intermittency \cite{nasuno_time-resolved_1998}, hysteresis \cite{degiuli_friction_2017}, protocol-dependence \cite{bililign_protocol_2019}, and the relaxation into limit cycles in which memories can be stored \cite{keim2018memory}. Each of these dynamics corresponds to different types of trajectories on the landscape of valid states. 

\section{Case studies}
\paragraph{Fragile states:} Earth landscapes are driven, out-of-equilibrium systems. Consider a mountain range: the horizontal convergence of tectonic plates leads to a piling up of rock to a critical angle, beyond which the wedge grows while maintaining a constant angle \cite{davis1983mechanics}. Locally, this growth is an effective topographic source term called uplift, which creates potential energy for transport. Erosion results as physical and chemical weathering break down rock into particulate material --- boulders down to soil --- which moves downhill by geophysical flows described above. Over geologic timescales a steady state is reached where erosion balances uplift on average \cite{anderson2010mechanics}. Because sediment transport rate increases rapidly for stresses above the yield point, hillsides \cite{ferdowsi2018glassy} and river channels \cite{phillips2016self} organize themselves, like a sandpile, to be in the vicinity of yield. This means that mountain landscapes flicker back and forth across the yield transition, due to environmental perturbations such rainfall/floods, freeze-thaw cycles, and earthquakes. 
Continental shelf environments are similar: sediment sourced from rivers and delivered to the shelf edge piles up on the continental slope, where earthquakes and storms trigger dense (debris) flows and dilute particulate (turbidity) currents that relax the over-steepened slope \cite{canals2004slope}. Thus, landscapes are driven to the fragile state --- where dynamics such as sub-yield creep, aging, hysteresis and failure occur. It is interesting that these behaviors, which are typically associated with glasses, occur in granular materials despite the absence of thermal energy. This suggests that mechanical noise may play a role akin to thermal fluctuations \cite{falk2011deformation, bandi2013fragility, reddy2011evidence, pons2016spatial, nicolas_deformation_2018}; we pick up this thread later.

\paragraph{Landscape patterns:}
 
In his seminal 1941 book \cite{Bagnold1941}, Bagnold laid out an approach for connecting the grain-scale physics of sand transport to the formation and evolution of wind-blown dunes. In the last two decades, a mature body of work by physicists has developed around testing and elaborating on his hypotheses. A key early result was the demonstration that the {\it saturation length} determines the length-scale of dune formation \cite{hersen2002relevant}. The saturation length is the distance needed to achieve balance between grain inertia and wind strength \cite{andreotti2010measurements}. This finding opened up sand dunes to laboratory exploration, because the density difference between water and air allowed the creation of scaled-down dunes underwater \cite{hersen2002relevant, reffet2010formation}. Models that include simplified aerodynamics, avalanching, and the saturation length were able to reproduce the pattern and scale of sand dunes observed in laboratory experiments \cite{reffet2010formation} and the field \cite{hersen2004corridors, schwammle2005model, ping2014emergence, zhang2010morphodynamics}. Even the feedbacks between vegetation growth and inhibition of sand transport have been encoded into models \cite{duran2006vegetation} that have been quantitatively confirmed with field data \cite{reitz2010barchan} (Box 2). 

Grain-scale sediment transport has also been connected to river-channel formation and associated landscape patterns. The simplest model for the cross-sectional shape of a river --- that the river-bed surface is at the threshold of motion --- has been quantitatively confirmed in the laboratory \cite{seizilles2013width, Reitz2014}. Remarkably, compilations of field data also show that the central tendency of natural rivers conforms to this prediction \cite{phillips2016self, metivier2017laboratory, dunne2018evidence}. Alluvial fans are cones of sediment built by a migrating river channel; experiments \cite{reitz2012experimental, delorme2018growth} and field observations \cite{parker1998alluvialII, miller2014generalized} have shown how the threshold of motion determines the overall shape of fan profiles (Box 2). Experiments and field observations of drainage network patterns, and their temporal evolution, show surprisingly good agreement with a theory for growth of threshold channels in a Laplacian field \cite{devauchelle2012ramification, berhanu2012shape}. This theory also reveals the connections of river network growth to a broader class of geophysical patterns that includes fracturing \cite{devauchelle2017laplacian}

\paragraph{Unifying fluid-sheared sediment transport:}
Particulate transport in rivers is traditionally separated into two regimes: (i) bed-load transport, in which particles move in close contact with and supported by the sediment bed; and (ii) suspension, where fluid-induced dispersion counteracts particle settling \cite{raudkivi1998loose} (Fig. \ref{fig:classify}). Moreover, wind-blown sand transport is distinguished from river flows by the larger significance of granular impacts on the entrainment of particles \cite{Bagnold1941} (Fig. \ref{fig:geo_materials}). A new and rapidly-developing understanding is emerging from explicit examination of granular dynamics, however, that is leading to a unified description of fluid-driven particulate flows. Laboratory experiments in laminar \cite{aussillous2013investigation, houssais2016rheology} and transitionally-turbulent \cite{allen2017depth} fluid flows have examined sediment transport by tracking particle motion from the interface to deep beneath, using refractive-index matched scanning. Results show three distinct regimes (Fig. \ref{fig:classify}a). The upper regime is a dilute granular suspension with low $\phi$ and large particle velocity $u$ that is dominated by hydrodynamic effects; at its base, $\phi$ rapidly increases and $u$ rapidly decreases at what may be considered the fluid-particle interface. The flowing layer of mobile grains below is the second, bed load, regime, characterized by approximately constant (and large) $\phi$ and exponentially decreasing $u$; granular frictional effects dominate here. Below the bed-load layer, a kink in the velocity profile to a slower, second exponential decay marks the transition to the third, creep, regime (Fig. \ref{fig:phase}). The entire range of bed-load to suspended-sediment transport, i.e., the first and second regimes, follows the $\mu(I)$ rheology \cite{houssais2016rheology, houssais2017toward}. Discrete Element Method (DEM) simulations driven by a mean-field fluid model have confirmed the applicability of $\mu(I)$ rheology to the turbulent flow regime \cite{maurin2016dense}. Such models have also demonstrated the importance of granular collision and viscous dissipation in determining the momentum balance, and resultant transport rate, of the flowing granular layer. These effects may be accounted for by introduction of a Stokes-like number \cite{duran2012numerical, pahtz2017fluid} which, together with $\mu(I)$, provides a unified framework for describing sediment transport by wind and water. Some predictions emerging from these grain-scale models have been confirmed by field measurements of sediment transport in rivers \cite{turowski2011start, phillips2014dynamics} and sand dunes \cite{martin2017wind} (Box 2).

\paragraph{Creep and the onset of flow:}
For almost a century, a simple Coulomb friction criterion has formed the basis for predicting the particle entrainment threshold by water \cite{shields1936anwendung} and wind \cite{Bagnold1941} flows. Recent Discrete Element Method (DEM) simulations, introduced above, have revealed important new insights: first, the importance of granular collision and viscous dissipation in determining the conditions for sustained transport \cite{duran2012numerical, pahtz2017fluid}; and second, the role of granular structure in modulating the local stability of the sediment bed \cite{clark2017role}. Complementary laboratory flume experiments, spanning the laminar to turbulent and low-to-high Stokes-number regimes, have confirmed the importance of these two factors \cite{charru2004erosion, frey2011bedload, allen2017depth, masteller2017interplay, lee2018determining}; they have also revealed the presence of creep below the onset of transport \cite{Houssais2015, allen2018granular}. This creep was found to strain harden the sediment bed \cite{masteller2017interplay, allen2018granular} and drive slow granular segregation \cite{ferdowsi2017river}. These dynamics are expected to produce hysteresis in the onset and cessation of transport, an effect that has been observed in natural rivers \cite{turowski2011start} (Box 2).

The failure and fluidization of landslides has also traditionally been described with a Coulomb model \cite{Terzaghi1943, schofield1968critical}. An important recent conceptual advance is the mapping of both landslide failure, and the onset of fluid-sheared sediment transport, to a creep-flow transition. In laminar flow experiments for the latter \cite{houssais2016rheology}, discussed above, transport occurred as creep below a critical viscous number $I_v \sim 10^{-5}$ (Fig. \ref{fig:phase}). 
Creep was characterized by (i) intermittent and localized particle rearrangements, that bear qualitative similarity to shear transformation zones in amorphous solids \cite{falk2011deformation}, and (ii) a departure from the expected $\mu(I)$ curve. At low enough driving stresses, creep occurred throughout the pack; above a critical stress, a flowing surface layer developed that was underlain by creep. Gravity-driven heap-flow experiments have exhibited all of the same creep behaviors \cite{amon2013experimental, komatsu2001creep, crassous2008experimental}, and DEM simulations have found a creep-flow transition at a critical inertial number $I_i \sim 10^{-5}$ \cite{ferdowsi2018glassy} --- in quantitative agreement with fluid-driven transport. The relation between stress and strain rate across the creep-flow transition in the simulations was consistent with a plastic depinning model recently proposed to describe yielding in glasses \cite{nicolas_deformation_2018, ferdowsi2018glassy}. Simulations have also added random disturbances to grain motion, meant to represent environmental disturbances in the field, and found that this influenced the rate but not the form of creep \cite{bendror2018controls, ferdowsi2018glassy}. Field measurements of creeping and fast landslides were found to be in fair agreement with simulations, indicating that important components of the creep-landslide transition are controlled by granular friction \cite{ferdowsi2018glassy}. Other field studies indicate that rate-weakening of accelerating landslides is common \cite{lucas2014frictional, handwerger2016rate, handwerger2019landslide} (Box 2); such behavior is likely granular in origin, but this link has not yet been made. As an interesting aside, experiments examining a dilute surface layer of bed-load transport found that the spatial patterning of mobile regions was consistent with a plastic depinning behavior \cite{aussillous2016scale}. Whether this transition on the \textit{surface} of a fluid-driven particle flow is related to the transition in the \textit{bulk} of a gravity-driven heap flow is unknown; however, both involve components of disorder and cooperative particle motion. The case of a dilute bed-load layer moving over a (quasi-)static bed is fascinating to consider for another reason: the energy landscape and the real (topographic) landscape are the same (Fig. \ref{fig:landscape}). Particles move over and around a disordered array of potential wells and barriers, but this landscape also evolves as particles are entrained from and deposited on the interface.

\section{Outstanding problems}

\paragraph{Athermal creep and the role of mechanical noise:} 
The sudden collapse and liquefaction of apparently solid soil, to form landslides and debris flows, is perhaps the most dramatic illustration of the need to better understand and predict the solid-liquid transition in granular materials. The best-studied scenario is landsliding induced by rainfall, which is invariably shown to enhance pore pressure; this effect has been presumed to drive the soil to yield \cite{roering2009using, handwerger2016rate, DiMaio2016, Lollino2017, handwerger2019landslide}. Earthquakes are another common driver of liquefaction \cite{meunier2008topographic}; even here, the mechanism typically invoked is shear-induced elevation of pore pressure \cite{sassa1996earthquake}. The most well-developed continuum models for landslide failure are built on two basic tenets from {\it critical state soil mechanics} \cite{schofield1968critical}: (i) a Mohr-Coulomb yield criterion, and (ii) pore-fluid pressure reduces contact forces by reducing the effective normal stress \cite{iverson1997debris, iverson2014depth}. Some issues with (i) were already discussed above; recent simulations have shown that the Mohr-Coulomb criterion fails even in weakly disordered materials, where the failure plane instead emerges from the coalescence of interacting damaged clusters \cite{dansereaucollective2019} (Fig. \ref{fig:classify}c). As for (ii) it was already suggested that lubrication, rather than pore pressure directly, may be the primary driver of liquefaction in geophysical flows \cite{ancey2001role}. If loss of rigidity arises from a frictional to lubrication transition, it would suggest in those cases that soil liquefaction is the mirror process of {\it discontinuous shear thickening} \cite{bonn2017yield, morris2018lubricated}. In some cases such as failure of muddy material underwater, lubrication may instead be localized at a basal slip surface leading to hydroplaning  \cite{mohrig1998hydroplaning, mohrig1999experiments, ilstad2004subaqueous}.

Any description of the solid-liquid transition in amorphous materials must account for creep; yet, creep in athermal granular systems is a frontier topic in soft-matter physics. The major challenge is that, unlike molecular glasses, concepts of `temperature' and `energy landscape' are poorly defined \cite{nicolas_deformation_2018}. Granular creep is generally understood to be a transient relaxation process that decays logarithmically with time \cite{hartley_logarithmic_2003}, as particles settle into more stable configurations. Yet, Earth materials creep indefinitely, presumably because of ceaseless mechanical disturbances. Besides pore pressure (rainfall) and shaking (earthquakes), geologists have invoked the generation of pore volume by bio-physical disturbances such as trees and animals \cite{culling1963soil, roering2004soil, roering2008well} to explain the creep of soil below the apparent angle of repose. These disturbances that are internal to the soil, rather than imposed at the boundaries, are evocative of thermal effects in glasses; but the applicability of thermal activation concepts to mechanical noise is currently an open question \cite{nicolas_deformation_2018} (Fig. \ref{fig:landscape}). One proposed way to conceptualize mechanical noise is as a stress that tilts the energy landscape (Fig. \ref{fig:landscape}), allowing particles to access a previously forbidden configuration --- typically through localized plastic rearrangments. In contrast to thermal systems, however, the energy landscape changes as particles rearrange \cite{agoritsas2015relevance, nicolas_deformation_2018}. Recent experiments/simulations have begun to explore the consequences of a range of disturbances on athermal creep. Acoustic driving \cite{johnson2008effects, ferdowsi2015acoustically} and vibrations \cite{griffa2011vibration} enhanced micro-slip and creep rates in confined granular systems. Intriguingly, acoustic \textit{emissions from} creeping \cite{johnson2013acoustic} and also fast-flowing \cite{elst2012auto} grains have been observed, which are now being related to stability and vibrational modes \cite{brzinski2018sounds}. This raises the tantalizing possibility of detecting precursor creep events on approach to failure using seismology in the field; though such applications are likely a long way off. Experiments with small stress modulations, imposed on a granular pack by an intruder, were able to induce a steady-state and effectively visco-elastic creep regime \cite{pons2016spatial}. Heap-flow DEM simulations in a channel also exhibited apparently steady-state creep, where the only imposed disturbance was the presence of walls \cite{ferdowsi2018glassy}. The upshot is this: although no formal theory mapping thermal to mechanical noise exists, the emerging phenomenological picture of athermal creep is that of glassy dynamics \cite{falk2011deformation, amon2013experimental, pons2016spatial}. In particular, (granular-friction mediated) relaxation and (mechanically-induced) rejuvenation drive persistent creep. Indeed, {\it creeping avalanches} observed in a thermally-influenced heap flow of micron-scale grains \cite{berut2017creeping} bear striking similarity to shear-localized rearrangements in a creeping heap flow of sand \cite{amon2013experimental}. Formalizing these similarities, and probing a wider variety of mechanical disturbances that are relevant to geophysical flows on land and undersea \cite{iverson1997debris, mohrig1999experiments, ilstad2004subaqueous, you2012dynamics, breard2016coupling}, are exciting challenges.

\paragraph{Active, and activated, matter:}
Landscape patterns on the Earth's surface are buffeted by a wide spectrum of forcings, from the scales of turbulent wind and water fluctuations to the fits and starts of plate tectonic motions. At first blush it is not obvious that steady-state landforms should exist at all. It turns out that the evolution of landscapes such as rivers and hillslopes to the (near-)critical state acts to filter out a wide range of environmental forcings \cite{jerolmack2010shredding, phillips2016self}, allowing the application of mean-field models for the driving stress. Soft-matter effects such as aging, hysteresis and multiple-stable states, however, suggest there are situations where mean-field approaches may fail. As a simple example, consider the consolidation of mud by dewatering. Sedimentation of clay particles forms aggregates \cite{allain1995aggregation} and colloidal gels \cite{mcanally2007management} with a microstructure reminiscent of a house of cards (Fig. \ref{fig:geo_materials}). Continued sedimentation induces an irreversible collapse under the hydrostatic burden \cite{barden1973collapse}, however, to produce a dense fabric of aligned clay particles with massively enhanced rigidity \cite{delage1984study}. Another example is the role of transient hydrodynamic forcing, such as the evaporation of suspensions that gives rise to colloidal films, cracks, and the celebrated coffee ring effect \cite{deegan1997capillary, Deegan2000}. Particles may be bonded by van der Waals, and even sintered, by capillary forces. Re-wetting does not restore the original suspension \cite{goehring2010evolution}, meaning that a time-averaged description of water content would not predict the state of matter. Fluctuating environmental forces on the Earth's surface are activating a range of mechanical responses that, ultimately, control the rigidity of soil and sediment in ways we have barely begun to explore. 

{\it Active matter,} in which particles move and/or exert forces, is now a firmly established research area in soft-matter physics \cite{marchetti_hydrodynamics_2013}. Yet, only recently have researchers explicitly shown that active matter can change the rheology and state transitions in glassy and granular materials  \cite{berthier_how_2017,junot_active_2017, saintillan2018rheology}. One study revealed how the presence of bacteria, even in modest concentrations, acts to suppress sedimentation of passive particles \cite{singh2017sedimentation}. This should be significant for muddy suspensions in bacteria-rich natural rivers and estuaries. In Earth-surface materials more broadly, active matter is pervasive; witness the bioturbation of mud, soil and gravel-river beds by innumerable organisms --- from worms to salmon to wombats \cite{butler1995zoogeomorphology, hassan2008salmon, wilkinson2009breaking, reinhardt2010dynamic}. Besides affecting transport rates, plants have been shown to qualitatively change river \cite{tal2007dynamic} and dune \cite{duran2006vegetation, reitz2010barchan} patterns. Geologists have awakened to the importance, and in some cases perhaps dominance, of biophysical processes in shaping the Earth's landsapes. Models developed to account for the effects of biota, however, are not based in mechanics; there is typically no explicit consideration of forces. In short, the Earth-surface is full of active matter, but active-matter approaches are absent. Small-scale physics experiments suggest some immediate avenues for exploration. One connection could be to link root growth into grains \cite{kolb_radial_2012,wendell_experimental_2012} to the mechanical wedging of tree roots that dilates soil and breaks down rock \cite{roering2004soil, roering2008well}. Insights from fiber-reinforced granular materials \cite{diambra_fibre_2010, dos_santos_mechanics_2010} may help us to think more mechanistically about root-reinforced hillslopes. Perhaps more distant but more intriguing: is pervasive bio-activation of soil effectively a creep rejuvation process that simply speeds up rates by tilting the energy landscape (Fig. \ref{fig:landscape}); or, does it produce a behavior that is mechanically distinct from the granular creep that we have encountered thus far?

\paragraph{Rheology of heterogeneous soft matter:} 
The rheology of geophysical flows is sensitive to particle size distribution and solids content \cite{major1992debris, coussot1994behavior, coussot1996recognition}. Consider again debris flows: slurries typically consisting of clay- to sand-sized particles and water, capable of entraining boulders. Increasing sand content has been shown to increase the yield stress \cite{scotto2010experimental}, and can even change bulk rheology from shear-thinning to shear-thickening \cite{bardou2007properties} (Box 1). We may speculate that the latter is related to the shutting off of lubrication associated with discontinuous shear thickening; however, it may also be due to large particles breaking up cohesive contact networks of clays. In debris flows, even subtle changes in rheology strongly influence strain-rate localization and boundary shear, and can lead to segregation of phases such as the formation of a granular-frictional front \cite{parsons2001experimental, leonardi2015granular}. The chemical properties of fine particles, especially surface charge, also matter. Different clay types produce varying suspension rheology that is dependent on salinity \cite{jeong2010rheological}, presumably due to cohesion. All of these factors influence the conditions for failure, and the destructive potential associated with runout, of debris flows. Considering failures underwater, the initial rheology of the grain mixture determines the degree of mixing with the overlying water, and can even switch  the failure mode from a gradually collapsing pile to a hydroplaning block \cite{mohrig1999experiments, ilstad2004subaqueous}.

These issues are at the forefront of soft-matter physics: what is the role of physical and chemical particle properties in the rheology and jamming of suspensions/granular flows?  The unifying framework of $\mu(I)$ rheology is appealing in its simplicity, and recent work has demonstrated how it may be generalized to account for: Non-Newtonian carrier fluids \cite{dagois2015rheology}; thermal effects \cite{wang2015constant}; and cohesion \cite{berger2016scaling, roy2017general}. On the other hand, qualitative changes in flow behavior may be induced by: particle polydispersity and shape \cite{nguyen_effects_2015, pednekar2018bidisperse}, surface roughness \cite{hsu2018roughness}, repulsion \cite{clavaud2017revealing} and hydrogen bonding \cite{james2019tuning}, attraction \cite{colombo2014stress}, and capillary forces \cite{koos2011capillary}. All of these factors ultimately influence particle microstructure, and explicit accounting for these changes in bulk continnum models is a challenge.
%

\section{Conclusions}

Landscapes are composed of, and formed by, flows of soft matter. By mapping the composition and dynamics of geophysical flows to recent advances in soft-matter physics, we hope to reveal the potential of the latter to help improve understanding of natural hazards and landscape evolution. In several cases of particulate-fluid flows examined here, this potential is already being realized. Soft matter approaches may be extended to other Earth materials. For example, solid rock \cite{li2011frictional} and ice \cite{goldsby2001superplastic} likely share much in common with amorphous solids such as glass --- albeit with additional complexities arising from partial melting and re-crystallization under high pressures --- while fragmented ice has been shown to behave as a jammed granular material \cite{burton_quantifying_2018}. Examining geophysical problems --- and their associated novel materials, geometries and boundary conditions --- can also reveal new physics or challenge existing frameworks. 
We see particular promise in building connections from grain to landscape scales, through the consideration of rheology, statistical physics, and athermal noise. 

\section{Acknowledgements}
The idea for this manuscript originated at the ``Physics of Dense Suspensions'' program at the Kavli Institute for Theoretical Physics, supported by the National Science Foundation (PHY-1748958). We are grateful to all participants of that workshop, especially the organizers: Bulbul Chakraborty, Emanuela Del Gado, and Jeff Morris. D.J.J. was sponsored by the Army Research Office (W911-NF-16-1-0290), the National Science Foundation (NRI INT 1734355), and the US National Institute of Environmental Health Sciences (P42ES02372).  K.E.D. is grateful for support from the National Science Foundation (DMR-1206808 and DMR-1608097) and the James S. McDonnell Foundation. We thank our research groups, and also Doug Durian and Paulo Arratia, for discussions that contributed to ideas presented here; and we thank Andrew Gunn for creating Figure \ref{fig:classify}. 

\section{Competing Interests Statement}
The authors declare that we have no competing interests.
\newpage 
\appendix
\renewcommand{\theequation}{\thesection.\arabic{equation}}

\noindent {\itshape This section should appear as a box within the paper}

\section{Box 1: Rheology of soft materials and Earth materials}

The most generic relation between shear stress $\tau$ and strain rate $ \dot{\gamma} $ for fluids is the Herschel-Bulkley relation,
\begin{equation} \label{eq: h-b}
    \tau = \tau_y + C \dot{\gamma}^n ,
\end{equation}
where $\tau_y$ is the yield stress. For a Newtonian fluid $\tau_y = 0$ and $n=1$, in which case $C = \eta_f$ is viscosity. For shear thickening (thinning) fluids $n>1$ ($n<1$), the apparent viscosity increases (decreases) with shear rate. Equation \ref{eq: h-b} has been used to describe a wide range of soft materials due to its flexibility. The physical origins of the yield stress and the exponent $n$ vary widely among systems, however, and for the most part remain to be understood \cite{chen_rheology_2010, bonn2017yield}. Natural and experimental debris flows typically behave as shear-thinning, yield-stress fluids that have been fit with Eq. \ref{eq: h-b} \cite{coussot1994behavior}.

The addition of particles to a Newtonian fluid creates a suspension that can be modeled as a single-phase, non-Newtonian fluid at high particulate volume fraction $\phi$. Herschel-Bulkley may be nondimensionalized by a confining pressure (normal stress) $P_p$, which re-casts the relation in terms of friction $\tau/P_p \equiv \mu $ and a non-dimensional shear rate $I_v = \eta_f \dot{\gamma}/P_p$ that we recognize as the viscous number \cite{boyer2011unifying, guazzelli2018rheology}:
\begin{equation} \label{eq: NDh-b}
    \mu = \mu_s + I_v^n.  
\end{equation}
Note that the ratio of shear to normal stresses at yield appears as a static friction coefficient, $\tau_y/P_p \equiv \mu_s $; but from the perspective of yield-stress fluids, this arises from a cooperative effect of many particles. For viscous (non-inertial) granular (athermal) suspensions it has been proposed that the effective friction is a result of two timescales; $t_\mathrm{micro} = \eta_f/P_p$ is a viscous drag timescale for a suspended particle, and $t_\mathrm{macro} = 1/\dot{\gamma}$ is the strain timescale for rearrangement of grains around a particle \cite{boyer2011unifying}. Accordingly, the constitutive relations for shear stress and volume fraction become functions of $I_v = t_\mathrm{micro}/t_\mathrm{macro}$:
\begin{equation} \label{eq: muIv}
    \tau = \mu (I_v) P_p \text{  and  } \phi = \phi (I_v),
\end{equation}
 Functional forms have been derived for Eq. \ref{eq: NDh-b} and Eq. \ref{eq: muIv}, and shown to fit a wide range of viscous granular suspensions \cite{boyer2011unifying, guazzelli2018rheology}. This rheology has been extended to sedimenting grains, and found to accurately describe the dense to dilute regimes of fluid-driven sediment transport \cite{houssais2016rheology} (see text). For flows where collisions dominate over fluid viscosity, the strain timescale remains the same but the relevant microscopic timescale for grain motion is inertial, $t_\mathrm{micro} = \sqrt{d^2 \rho_p /P_p}$ where $d$ and $\rho_p$ are particle diameter and density, respectively. Different functional forms for Eq. \ref{eq: NDh-b} and Eq. \ref{eq: muIv}, with an inertial number $I_i$ in place of $I_v$, are found to describe a wide range of inertial granular flows \cite{midi2004dense} and also natural landslides \cite{ferdowsi2018glassy}. Importantly, functional relations based on Eq. \ref{eq: muIv} all exhibit an effective friction that converges to the static value in the limit of vanishing shear rate. This yield transition is associated with a packing fraction that approaches the critical value $\phi_c$ associated with jamming  \cite{liu_jamming_2010, boyer2011unifying}. These relations are collectively referred to as $\mu(I)$ rheology.

In repulsive colloidal suspensions, the excluded-volume effects that dominate $\mu(I)$ rheology at high-$\phi$ values are still relevant. Thermal affects introduce an additional relaxation timescale, however, such that high-$\phi$ {\it colloidal glasses} are typically considered to be distinct from granular systems in terms of yielding \cite{chen_rheology_2010, bonn2017yield}. Nonetheless, simulations have shown that $\mu(I)$ may be generalized to repulsive colloidal glasses by explicitly accounting for thermal effects via a Peclet number \cite{wang2015constant}.

Other classes of soft matter may be created by combinations of the above classes. One relevant example for geophysical flows is granular suspensions in yield stress fluids, recently examined by the addition of repulsive and non-Brownian particles to non-Newtonian emulsions \cite{dagois2015rheology}. Volume exclusion effects influence the yield stress and effective viscosity, independent of the suspending fluid composition. Accordingly, Eq.  \ref{eq: h-b} may be generalized to:
\begin{equation} \label{eq: h-b-phi}
    \tau = \tau_{y, \phi}(\phi)\tau_y + C_{\phi}(\phi)C \dot{\gamma}^n ,
\end{equation}
where $\tau_y$, $C$ and $n$ are properties of the suspending fluid, and $\tau_{y, \phi}(\phi)$ and $C_{\phi}(\phi)$ are dimensionless functions that increase motonically from 1 with increasing $\phi$. As with Eq. \ref{eq: h-b}, Eq. \ref{eq: h-b-phi} may be recast equivalently in terms of $\mu(I)$ rheology \cite{dagois2015rheology}. This model has relevance for natural debris flows, which often consist of dense mud suspensions that carry boulders.

\newpage 
\noindent {\itshape This section should appear as a box within the paper}

\section{Box 2: Geophysical field methods for soft landscapes}

{\bf Sediment transport:}
Novel geophysical methods are allowing us to probe the mechanics and dynamics of fluid-driven sediment transport in field settings. The dispersion of radio-tagged cobbles in rivers \cite{hassan2016coarse} has been used to validate bed-load transport models \cite{phillips2014dynamics}, and smart rocks are being developed that can actually measure the forces of grain collisions \cite{underwood2012build}. Sediment transport rates in rivers and dune fields have also been estimated from a wide range of techniques; for example, impact plates \cite{turowski2011start}, optical gates \cite{martin2017wind}, and passive \cite{geay2017passive} and active \cite{church2012gravel} acoustics. 

{\bf Flow fields:}
The data required to determine relevant weather conditions for wind-blown dunes \cite{horel2002mesowest}, and the hydrology for river channels \cite{lins2012usgs}, are freely available for many locations in the USA. This information provides the boundary conditions for flows impinging on sediment beds, and allows estimation of the time-averaged fluid shear stresses that landforms adjust to \cite{phillips2016self}. More detailed measurements are needed to critically test sediment transport models. Fluid velocity and turbulent (Reynolds) stress profiles are now routinely collected in rivers and atmospheric flows using acoustic and optical doppler techniques \cite{church2012gravel, martin2017wind, de2018study}. Many of these methods are also deployed in the laboratory; for example, acoustic techniques are often applied to optically-opaque particulate suspensions such as turbidity currents, in order to image the internal structure of these flows \cite{perillo2015acoustic}.

{\bf Topography:}
The explosion of high-resolution topographic field data has transformed the discipline of geomorphology. In particular, ground-based \cite{telling2017review} and aerial LiDAR topography (Light Detection And Ranging) are rendering high-fidelity digital models of terrestrial landscapes that facilitate stringent hypothesis testing. These datasets are rapidly expanding in number and global coverage, and many are freely available online \cite{krishnan2011opentopography}. Access to seafloor topographic data (bathymetry), collected from seismic surveys conducted by boat, is also growing quickly \cite{john2018boem}. When coupled with mass conservation and some knowledge of boundary conditions, topography may be used to assess the rheological behavior of Earth materials over geologic time \cite{roering2004soil, ferdowsi2018glassy}. In fast-changing landscapes such as active landslides \cite{roering2009using} and migrating sand dunes \cite{reitz2010barchan}, repeat topographic surveys have been used to directly measure spatial patterns and rates of erosion and deposition. Moreover, expanding data coverage facilitates the exploration and discovery of fascinating new Earth--surface patterns.

{\bf Slipping:}
Geotechnical measurements of active landslides in the field can produce highly-resolved ground displacements that constrain the kinetics. Vertical velocity profiles in soil are often collected within boreholes, which measure the angular displacements of a string of inclinometer sensors  \cite{Yufei}. Slope movement is often driven by fluctuations in groundwater levels, so some studies also collect precipitation and water table measurements \cite{Lollino2017}. Vertical deformation and soil moisture profiles may alternatively be collected with Time-Domain Reflectometry, an impedance technique \cite{Lin2009}. Spatially-extended data on surface-soil motion is also collected using GPS \cite{Wang2011} and Interferometric Synthetic Aperture Radar (InSAR) \cite{roering2009using, Zaugg2016}; these data complement the depth-resolved, but spatially-localized, deformation profiles from boreholes. InSAR was recently used to document, in stunning spatial and temporal resolution, the creep to landslide transition on a California mountain side \cite{handwerger2019landslide}. These field measurements are often coupled with laboratory tests of soil mechanical properties, using samples extracted from the field \cite{DiMaio2016}. Machine-learning algorithms have been applied to ground-displacement and associated environmental data for hillslopes, in hopes of enhancing forecasting of landslides in a fully automated and cost-effective manner \cite{Tordesillas2018}. 

{\bf Seismology:}
A rapidly developing area is {\it seismic geomorphology,} which capitalizes on decades of advances in seismology (motivated by earthquakes), and the wide distribution of seismic arrays, to determine the location and magnitude of sediment transport \cite{burtin2016seismic, roth2016bed}. Passive seismic monitoring has been used to detect rigidity changes in soil preceding landslide failure \cite{mainsant2012ambient} and to interpret flow dynamics \cite{bertello2018dynamics}. The seismic noise resulting from transport in the field shares tantalizing similarities with acoustic emissions from failing  materials in the laboratory \cite{garcimartin_statistical_1997, johnson2013acoustic, brzinski2018sounds}; this should be further explored. Finally, seismic geomorphology has also examined how landscapes respond to shear imposed by earthquakes \cite{burtin2016seismic}.

\newpage 


\end{document}